\documentclass[10pt]{bmc_article}    

\usepackage{cite} 
\usepackage{url}  
\usepackage{ifthen}  
\usepackage{multicol}  
\usepackage[utf8]{inputenc}
\usepackage{graphicx}
\usepackage{float}
\usepackage{geometry}
\usepackage{epstopdf}
\usepackage[T1]{fontenc}
\usepackage{color}
\usepackage{marvosym}
\usepackage{colortbl}
\usepackage{longtable}
\usepackage{amsmath}
\usepackage{amssymb}
\usepackage{amsthm}
\usepackage{multirow}
\usepackage{rotating}
\usepackage{caption}

\usepackage[nofiglist,notablist]{endfloat}

\DeclareSymbolFont{letters}{OML}{ztmcm}{m}{it}
\DeclareSymbolFontAlphabet{\mathnormal}{letters}

\newboolean{publ}

\newenvironment{bmcformat}{\begin{raggedright}\baselineskip20pt\sloppy\setboolean{publ}{false}}{\end{raggedright}\baselineskip20pt\sloppy}

\def \bussname {$\pi$BUSS}

\begin{document}
\begin{bmcformat}

\title{\bussname: a parallel BEAST/BEAGLE utility for sequence simulation under complex evolutionary scenarios
}

\author{Filip Bielejec\correspondingauthor$^{1}$%
       \email{Filip Bielejec\correspondingauthor - filip.bielejec(at)rega.kuleuven.be}%
      \and
         Philippe Lemey$^{1}$%
     \and
         Luiz Max Carvalho$^{2}$%
     \and
         Guy Baele$^{1}$%
      \and
         Andrew Rambaut$^{3}$ %
     and
         Marc A.~Suchard$^{4,5}$
      }
      
\address{%
    \iid(1)Department of Microbiology and Immunology, Rega Institute, KU Leuven, Leuven, Belgium\\
    \iid(2)Program for Scientific Computing (PROCC), Funda\c{c}\~ ao Oswaldo Cruz, Rio de Janeiro, Brazil \\
    \iid(3)Institute of Evolutionary Biology, University of Edinburgh, Edinburgh, United Kingdom\\
    \iid(4)Departments of Biomathematics and Human Genetics, David Geffen School of Medicine at UCLA, University of California, Los Angeles, CA, 90095, USA\\
    \iid(5)Department of Biostatistics, UCLA Fielding School of Public Health, University of California, Los Angeles, CA, 90095, USA}%

\maketitle

\ifthenelse{\boolean{publ}}{\begin{multicols}{2}}{}

\begin{abstract}
        
\paragraph*{Background:} 
Simulated nucleotide or amino acid sequences are frequently used to assess the performance of phylogenetic reconstruction methods.
BEAST, a Bayesian statistical framework that focuses on reconstructing time-calibrated molecular evolutionary processes,
supports a wide array of evolutionary models, but lacked matching machinery 
for simulation of character evolution along phylogenies. 
      
\paragraph*{Results:} 
We present a flexible Monte Carlo simulation tool, called {\bussname}, that employs the BEAGLE high performance library for phylogenetic computations within BEAST to rapidly generate large sequence alignments under complex evolutionary models.
{\bussname} sports a user-friendly graphical user interface (GUI) that allows combining a rich array of models across an arbitrary number of partitions. 
A command-line interface mirrors the options available through the GUI and facilitates scripting in large-scale simulation studies.
Analogous to BEAST model and analysis setup, more advanced simulation options are supported through an extensible markup language (XML) specification, which in addition to generating sequence output, also allows users to combine simulation and analysis in a single BEAST run.

\paragraph*{Conclusions:} 
{\bussname} offers a unique combination of flexibility and ease-of-use for sequence simulation under realistic evolutionary scenarios. 
Through different interfaces, {\bussname} supports simulation studies ranging from modest endeavors for illustrative purposes  to complex and large-scale assessments of evolutionary inference procedures.
The software aims at implementing new models and data types that are continuously being developed as part of BEAST/BEAGLE.
 
\end{abstract}

\noindent (Keywords: Simulation, Phylogenetics, BEAST, BEAGLE, evolution, Monte Carlo
)\\


\section*{Background}

Recent decades have seen extensive development in phylogenetic inference, resulting in a myriad of techniques, each with specific properties concerning
evolutionary model complexity, inference procedures and performance both in terms of speed of execution and estimation accuracy.
With the development of such phylogenetic inference methods comes the need to synthesize evolutionary data in order to compare estimator performance and to characterize strengths and weaknesses of different approaches.
Whereas the true underlying evolutionary relationships between biological sequences are generally unknown,
Monte Carlo simulation allows generating test scenarios while controlling for the evolutionary history as well as the tempo and mode of evolution. 
This has been frequently used to compare the performance of tree topology estimation (e.g. \cite{stamatakis05}), but it also applies to evolutionary parameter estimation and ancestral reconstruction problems (e.g. \cite{blanchette08}).
In addition, Monte Carlo sequence simulation has proven useful for assessing model adequacy (e.g. \cite{brown2009}) and for testing competing evolutionary hypotheses (e.g. \cite{goldman93}).
It is therefore not surprising 
that several general sequence simulation programs have been developed (e.g. Seq-Gen \cite{rambaut1997seq}), but also inference packages that do not primarily focus on tree reconstruction, such as PAML \cite{PAML} and HyPhy \cite{HyPhy}, maintain code to simulate sequence data under the models they implement.

As a major application of phylogenetics, estimating divergence times from molecular sequences requires an assumption of roughly constant substitution rates throughout evolutionary history \cite{zuckerkandl62}. 
Despite the restrictive nature of this molecular clock assumption, its application in a phylogenetic context has profoundly influenced modern views on the timing of many important events in evolutionary history \cite{arbogast2002}. 
Following a long history of applying molecular clock models on fixed tree topologies, 
the Bayesian Evolutionary Analysis by Sampling Trees (BEAST) package \cite{BEAST}
fully integrates these models, including more realistic relaxed clock models \cite{Drummond2006,Drummond2010}, in a phylogenetic inference framework.
By focusing on time-measured evolutionary histories, BEAST has grown increasingly popular as a general framework for historical inference of evolutionary and population genetic processes.
The software integrates substitution models, molecular clock models, tree-generative (coalescent or birth-death) models and trait evolutionary models in a modular fashion, allowing the user to combine different parameterizations for each module.
The ability to perform joint inference over a flexible combination of models and data types overcomes the pitfall of traditional stepwise inference procedures that purge intermediate estimates (e.g. a tree topology) of their uncertainty, but greatly complicates performance assessment.
Here, we address this by introducing a parallel BEAST/BEAGLE utility for sequence simulation ({\bussname}).

\section*{Implementation}

We develop several sequence simulation procedures that balance between ease-of-use and accessibility to model complexity.
On the one hand, the core simulation routine is implemented in the BEAST software package \cite{BEAST} and simulations can be performed following specifications in an extensible markup language (XML) input file (Figure~\ref{fig:screenshot} A).
This procedure provides the most comprehensive access to the BEAST arsenal of models, but may require custom XML editing.
On the other hand, {\bussname} also represents a stand-alone package that conveniently wraps the simulation routines in a user friendly graphical user interface (GUI), allowing users to set up and run simulations by loading input, selecting models from drop-down lists, setting their parameter values, and generating output in different formats (Figure~\ref{fig:screenshot} B).
To facilitate scripting, {\bussname} is further accessible through a command-line interface (CLI), with options that mirror the GUI.
The simulation routines are implemented in Java and interface with the Broad-platform Evolutionary Analysis General Likelihood Evaluator (BEAGLE) high-performance library \cite{Ayres} through its application programming interface (API) for computationally intensive tasks. 

\begin{figure}[h]
\begin{center}
\includegraphics[scale=0.55]{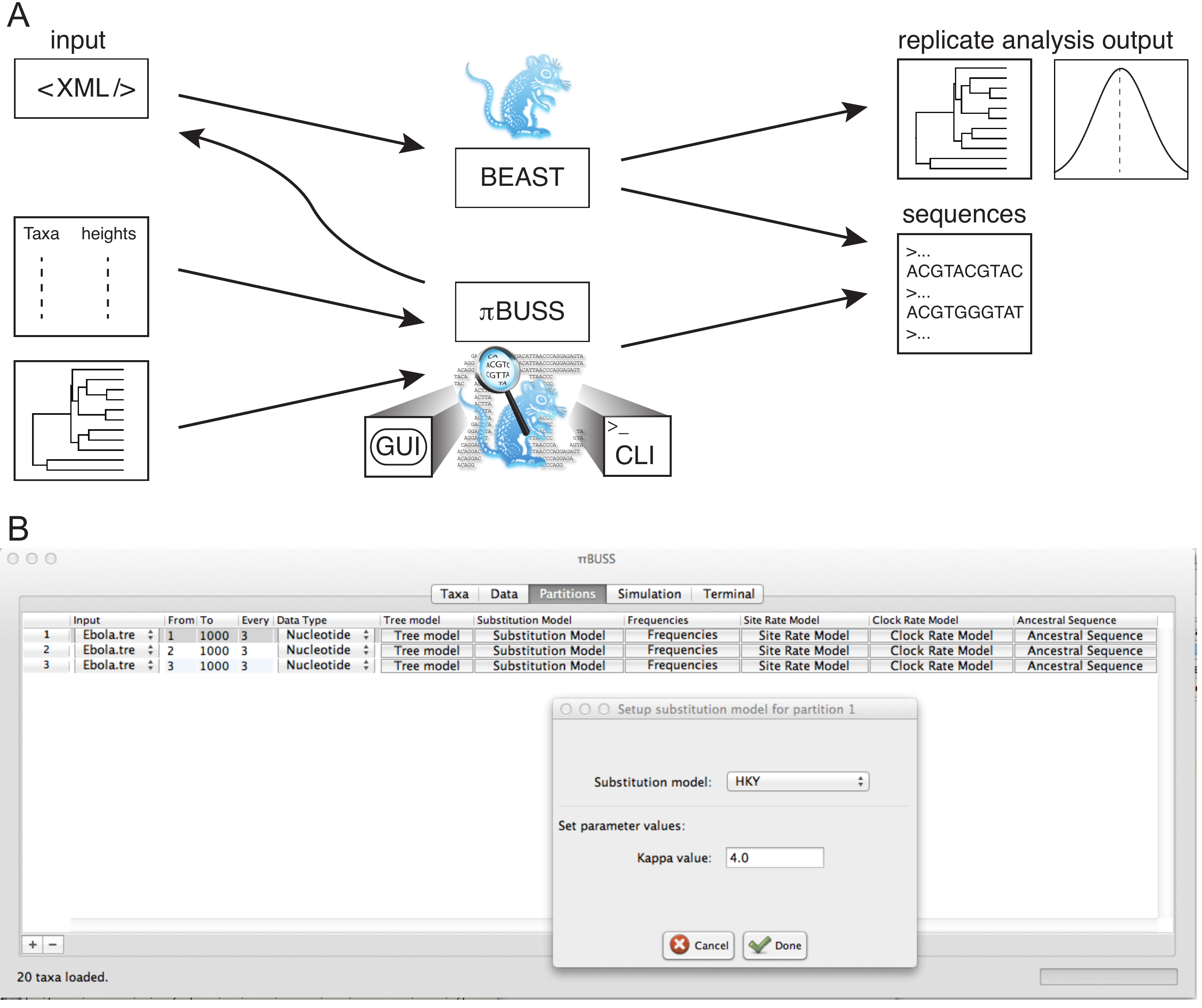} 
\end{center}
\caption{
{\bf Overview of the {\bussname} simulation procedures and GUI screenshot.}
A. Schematic representation of the different ways to employ the {\bussname} simulation software.
Based on an XML input file, simulations can be performed using the core implementation in BEAST.
BEAST can generate sequence data as well as analyze the replicate data in a single run.
Using both the GUI or CLI, {\bussname} can run simulations based on an input tree or a list of taxa and their heights.
The software can also write the simulation settings to an XML file for simulation in BEAST.
B. The screenshot example shows the set-up of a codon position partitioned simulation in the Partitions panel of the graphical user interface.
The Hasegawa, Kishino and Yano (HKY) model is being set as the substitution model for partition 1, with a $\kappa$ (the transition-transversion bias) parameter value of $4.0$.
}
\label{fig:screenshot}
\end{figure}

\subsection*{Program input} 
Similar to standard evolutionary analyses using BEAST, the core implementation of the software can be invoked by loading an XML file with simulation settings.
The simulation procedure requires a user-specified tree topology or a set of taxa with their heights (inversely proportional to their sampling time) for which a tree topology can be simulated using a coalescent model.
Setting all heights to 0 would be equivalent to contemporaneously-sampled taxa.
In {\bussname}, such a tree can be loaded in NEXUS or NEWICK format, or a taxa list can be set-up in the Data panel for subsequent coalescent simulation of the genealogy.
Creating the latter is further facilitated by the ability to load a tab-delimited file with a set of taxa and their corresponding heights.
The input tree or taxon list can also be specified through the command-line interface of {\bussname}.
 
\subsection*{Program output} 

{\bussname} generates sequence output in FASTA format but it also supports XML output of the simulation settings. 
Not only does the XML provide a record of the settings, but the file can be loaded in BEAST for sequence simulation.
In this sense, {\bussname} is analogous to the BEAUti tool that generates XML input for BEAST analysis \cite{BEAST}, and a {\bussname}-generated XML file specifying standard models can serve as a template for editing more complex simulations.
Moreover, the XML file can also be edited in order to directly analyze the generated sequence data, which avoids writing the sequence data to a file.

\subsection*{Models of evolution} 

{\bussname} is capable of generating trees from a list of taxa using simple coalescent models, including a constant population size or exponential growth model.
The software supports simulation of nucleotide, amino acid and codon data along the simulated or user-specified phylogeny using standard substitution models.
For nucleotide data, the Hasegawa, Kishino and Yano model (HKY; \cite{hky85}), the Tamura Nei model (TN93; \cite{TN93}) and the general time-reversible model (GTR; \cite{GTR}) can be selected from a drop-down list, and more restrictive continuous-time Markov chain (CTMC) models can be specified by tailoring parameters values.  
Coding sequences can be simulated using Goldman and Yang's model of codon evolution (GY94; \cite{Goldman1994}), which is parameterized in terms of a non-synonymous and synonymous substitution rate ratio ($dN/dS$ or $\omega$) and a  transition/transversion rate ratio ($\kappa$).
Several empirical amino acid substitution models are implemented, including the Dayhoff \cite{dayhoff}, JTT \cite{jtt}, BLOSUM62 \cite{blosum62}, WAG \cite{WAG} and LG \cite{LG} model.
Equilibrium frequencies can be specified for all substitution models as well as among-site rate heterogeneity through the widely-used discrete-gamma distribution \cite{Yang96} and proportion of invariant sites \cite{Gu01071995}. 

An important feature of {\bussname} is the ability to set up an arbitrary number of partitions for the sequence data and associate independent substitution models to them.
Such settings may reflect codon position-specific evolutionary patterns or approximate genome architecture with separate substitution patterns for coding and non-coding regions.
Partitions may also be set to evolve along different phylogenies, which could be used, for example, to investigate the impact of recombination or to assess the performance of recombination detection programs in specific cases.
Finally, partitions do not need to share the exact same taxa (e.g. reflecting differential taxon sampling), and in partitions where a particular 
taxon is not represented the relevant sequence will be padded with gaps. 

Inspired by the BEAST framework, {\bussname} is equipped with the ability to simulate evolutionary processes on trees calibrated in time units. 
Under the strict clock assumption, this is achieved by specifying an evolutionary rate parameter that scales each branch from time units into substitution units.
{\bussname} also supports branch-specific scalers drawn independently and identically from an underlying distribution (e.g. log normal or inverse Gaussian distributions), modeling an uncorrelated relaxed clock process \cite{Drummond2006}.
Simulations do not need to accommodate an explicit temporal dimension and input trees with branch lengths in substitution units will maintain these units with the default clock rate of 1 (substitution/per site/per time unit).

The data types and models described above are available through the {\bussname} GUI or CLI, but additional data types and more complex models can be specified directly in an XML file.
This allows, for example, simulating any discrete trait, e.g. representing phylogeographic locations, under reversible and nonreversible models \cite{Lemey2009,edwards2011ancient}, with potentially sparse CTMC matrices \cite{Lemey2009}, as well as simulating a combination of sequence data and such traits.
As an example of available model extensions is the ability to specify different CTMC matrices over different time intervals of the evolutionary history, allowing for example to model changing selective constraints through different codon model parameterizations or seasonal migration processes for viral phylogeographic traits \cite{bielejec2013}.


\section*{Results and Discussion}

We have developed a new simulation tool, called {\bussname}, that we consider to be a rejuvenation of Seq-Gen \cite{rambaut1997seq}, with several extensions to better integrate with the BEAST inference framework.
Compared to Seq-Gen and other simulation software (Table~\ref{tab:Sim}), {\bussname} covers a relatively wide range of models while, similar to Mesquite, offering a cross-platform, user-friendly GUI.
{\bussname} is implemented in the Java programming language, and therefore requires a Java runtime environment, and depends on the BEAGLE library.
Although speed is unlikely to be an impeding factor in some simulation 
efforts,
the core implementation using the BEAGLE library provides substantial increases in speed for large-scale simulations, in particular when invoking multi-core architecture to produce highly partitioned synthetic sequence data.

\begin{sidewaystable}[h]
\caption{{ \footnotesize {\bf Comparison between a selection of sequence simulation packages.} We compared the availability of evolutionary modeling options and software implementation aspects of eight commonly used sequence simulation packages. `X' indicates presence of the feature.
}}
\footnotesize{
\begin{tabular}{lccccccccccc}
\hline 
&\textbf{Feature/Package} &{\bussname} & Seq-Gen & PhyloSim & Recodon &Indelible \cite{indelible} & DAWG \cite{dawg} & Mesquite \cite{mesquite} & r8tes  \cite{r8tes} & Rose \cite{rose} & Evolver \cite{PAML} \tabularnewline
\hline
\multirow{6}{*}{Evolutionary modeling}&\textbf{Codon models}    &X &X & &X & & &X& & & X \tabularnewline
 &\textbf{Amino acid models} &X & & X &  &X & &X & &X &X \tabularnewline
 &\textbf{Indel models} & & X\footnote{Seq-Gen does not include indel simulation but the modified indel-Seq-Gen does \cite{indseqgen}. } & & X & & X&X & &X &\tabularnewline
 &\textbf{Partitions} &X & X & X & & X& &X & & &X\tabularnewline 
 &\textbf{Relaxed clock models} &X & & & & & & &X & & \tabularnewline 
 &\textbf{Ancestral sequences}\footnote{The ability to annotate ancestral characters at internal nodes.} &  &  & X & X &X & & & & & X\tabularnewline 
  &\textbf{Coalescent simulation} &X & & X & & &X &X & & &X \tabularnewline
 &\textbf{Recombination} & X & X & & X& & & & &\tabularnewline
 &                            &  &  & & & &\tabularnewline
 \multirow{5}{*}{Implementation}&\textbf{Graphical interface} &X &  & & & & &X & & &\tabularnewline
 &\textbf{Multicore}&X & & X & & & & & & &\tabularnewline
 &\textbf{Cross-platform}&X & X & X &  &X &X &X & &X &X\tabularnewline
\end{tabular}
} 
\label{tab:Sim}
\end{sidewaystable}

\subsection*{Program validation}

We validate {\bussname} in several ways. 
First, we compare the expected site probabilities, as calculated using tree pruning recursion \cite{Felsenstein1981}, with the observed counts resulting from {\bussname} simulations. 
To this purpose, we calculate the probabilities for all $4^3$ possible nucleotide site patterns observed at the tips of a particular 3-taxon topology using an HKY model with a discrete gamma distribution to model rate variation among sites.
We then compare these probabilities to long-run ($n = 100,000$) site pattern frequencies simulated under this model and observe good correspondence in distribution (Pearson's $\chi^2$ test, $p = 0.42$).

%

We also perform simulations over larger trees and estimate substitution parameters (e.g. $\kappa$ in the HKY model) using BEAST for a large number of replicates. 
Not only do the posterior mean estimates agree very well with the 
simulated values,
but we also find close to nominal coverage, and relatively small bias and variance (mean squared error).
These good performance measures have also recently been demonstrated for more complex substitution processes \cite{bielejec2013}.

%

\subsection*{Example application}

We illustrate the use of simulating sequence data along time-calibrated phylogenies to explore the limitations of estimating old divergence times for rapidly-evolving viruses.  
Wertheim and Kosakovsky Pond \cite{Wertheim2011} examine the evolutionary history of Ebola virus from sequences sampled over the span of three decades.
Although maintaining remarkable amino acid conservation, the authors estimate nucleotide substitution rates on the order of $10^{-3}$ substitutions/per site/per year and a correspondingly recent time to most recent common ancestor (tMRCA) of about 1,000 years.
These estimates suggest a strong action of purifying selection to preserve amino acid residues over longer evolutionary time scales, which may not be accommodated by standard nucleotide substitution models.
The authors demonstrate that accounting for variable selective pressure using codon models can result in substantially older origins in such cases.

Here, we explore the effect of temporally varying selection pressure throughout evolutionary history on estimates of the tMRCA using nucleotide substitution models.
In particular, we model a process that is characterized by increasingly stronger purifying selection as we go further back in to time. 
To this purpose, we set up an `epoch model' that specifies different GY94 codon substitution processes along the evolutionary history \cite{bielejec2013}, and parameterize them according to a log-linear relationship between time and $\omega$.
Specifically, we let the process transition from $\omega$ = $1.0$, $0.2$, $0.1$, $0.02$, $0.01$, $0.002$, and $0.001$ at time = $10$, $50$, $100$, $500$, $1000$ and $5000$ years in the past, respectively.
We simulate a constant population size genealogy of 50 taxa, sampled evenly during a time interval of 25 years, and simulate sequences according to the time-heterogeneous codon substitution process with a constant clock rate of $3 \times 10^{-3}$ codon substitutions/codon site/year.
By producing 100 replicates for various population sizes ($1$, $5$, $10$, $50$, $100$, $500$ and $1000$), the sequences are simulated over genealogies with different tMRCAs.
We note that under this model, trees with tMRCAs of about 10,000 years still result in sequences with a noticeable degree of homology (resulting in a mean amino acid distance of about 0.5, which is in the same range of the mean amino acid distance for sequences representative of the primate immunodeficiency virus diversity).
Using a constant $\omega$ of 0.5 on the other hand results in fairly randomized sequences.
We subsequently analyze the replicate data using a codon position partitioned nucleotide substitution model in BEAST and plot the correspondence between simulated and estimated tMRCAs in Figure \ref{fig:deep_root}.

\begin{figure}[h]
\begin{center}
\includegraphics[scale=1.0]{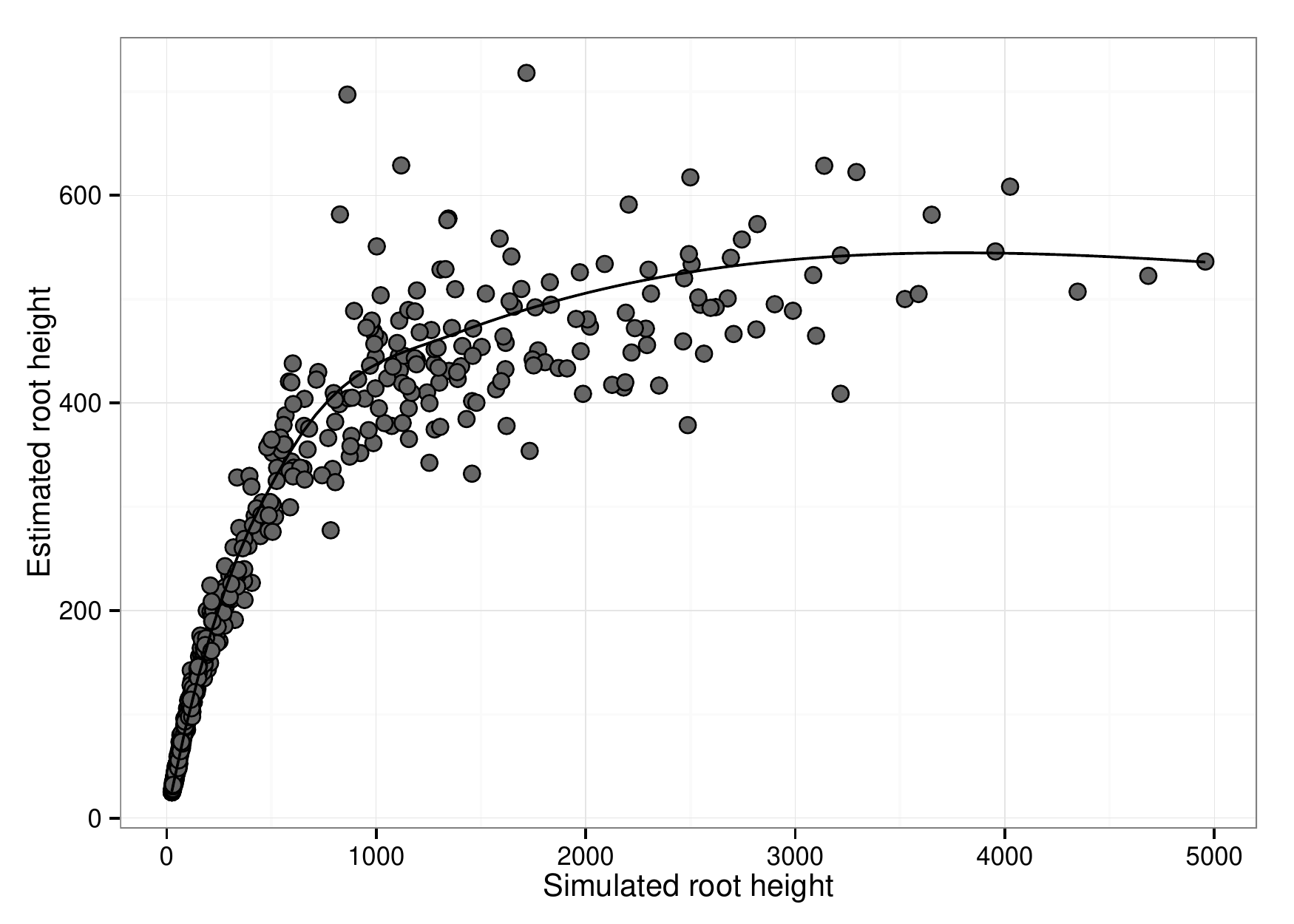} 
\end{center}
\caption{
{\bf Correspondence between simulated and estimated tMRCAs when purifying selection increases back in time in simulated data sets.} 
}
\label{fig:deep_root}
\end{figure}

Our simulation exercise shows that a linear relationship between simulated and estimated tMRCAs only holds for 100 to 200 years in the past, and estimates quickly level off after about 1000 years in the past.
This can be explained by the unaccounted decline in amino acid substitutions and saturation of the synonymous substitutions as we go further back in time.
Although we are not claiming that evolution occurs quantitatively or even qualitatively according to the particular process we simulate under, and we ignore other confounding factors (such as potential selective constraints on non-neutral synonymous sites), this simulation does conceptualize some of the limitations to estimating ancient origins for rapidly evolving viruses that experience strong purifying selection over longer evolutionary time scales.


\section*{Conclusion}

{\bussname} provides simulation procedures under many evolutionary models or combinations of models available in the BEAST framework.  
This feature facilitates the evaluation of estimator performance during the development of novel inference techniques and the generation of predictive distributions under a wide range of evolutionary scenarios that remain critical for testing competing evolutionary hypotheses.
Combinations of different evolutionary models can be accessed through a GUI or CLI, and further extensions can be specified in XML format with a syntax familiar to the BEAST user community.
Analogous to the continuing effort to support model set-up for BEAST in BEAUti, future releases of {\bussname} aim to provide simulation counterparts to the BEAST inference tools, both in terms of data types and models.
Interesting targets include discrete traits, which can already be simulated through XML specification, continuously-valued phenotype data \cite{Lemey2010} and indel models.
Finally, {\bussname} provides opportunities to pursue further computational efficiency through parallelization on advancing computing technology.
We therefore hope that {\bussname} will further stimulate the development of sequence and trait evolutionary models and contribute to advancement of our knowledge about evolutionary processes.

\section*{Availability and requirements}
The parallel BEAST/BEAGLE utility for sequence simulation is licensed under the GNU Lesser GPL and its source code is freely available as part of the BEAST GoogleCode repository:
\url{http://code.google.com/p/beast-mcmc/}

Compiled, runnable package targeting all major platforms along with a tutorial and supplementary data are hosted at:
\url{http://rega.kuleuven.be/cev/ecv/software/pibuss}.

{\bussname} requires Java runtime environment version 1.5 or later to run its executables and the BEAGLE library to be present and configured to fully utilize its capabilities.

\section*{Authors' contributions}
FB designed and implemented the software. 
Large portions of code extend or are based on BEAST interfaces designed and developed by AR and MAS. 
PL and GB conceived the original idea and helped with the design of the software. 
PL designed the simulation study employing the software.
LMC wrote the software tutorial, tested and commented on the software. 
All authors contributed to the writing of this manuscript.

\section*{Acknowledgements}
The research leading to these results has received funding from the European Research Council under the European Community's Seventh Framework Programme (FP7/2007-2013) under Grant Agreement no. 278433-PREDEMICS and ERC Grant agreement no. 260864, the Welcome Trust (grant no. 092807) and the National Institutes of Health (R01 GM086887 and R01 HG006139) and the National Science Foundation (IIS 1251151 and DMS 1264153).
The National Evolutionary Synthesis Center (NESCent) catalyzed this collaboration through a working group (NSF EF-0423641). LMC would like to thank the Program for Scientific Computing staff for operational support.


{\ifthenelse{\boolean{publ}}{\footnotesize}{\small}
\bibliographystyle{bmc_article}  
\bibliography{BussBMCBioinf} }     

\ifthenelse{\boolean{publ}}{\end{multicols}}{}


\end{bmcformat}
\end{document}